\documentclass[letterpaper]{article} 
\usepackage{aaai2026} 
\usepackage{times}  
\usepackage{helvet}  
\usepackage{courier}  
\usepackage[hyphens]{url}  
\usepackage{graphicx} 
\usepackage{natbib}  
\usepackage{caption} 
\usepackage{algorithm}
\usepackage{algorithmic}

\usepackage{newfloat}
\usepackage{listings}
\DeclareCaptionStyle{ruled}{labelfont=normalfont,labelsep=colon,strut=off} 
\floatstyle{ruled}
\newfloat{listing}{tb}{lst}{}
\floatname{listing}{Listing}
\title{Auto-BenchmarkCard: Automated Synthesis of Benchmark Documentation}
\author {
    Anonymous Submission
}

\author {
    Aris Hofmann\textsuperscript{\rm 1,\rm 2},
    Inge Vejsbjerg\textsuperscript{\rm 2},
    Dhaval Salwala\textsuperscript{\rm 2},
    Elizabeth M. Daly\textsuperscript{\rm 2}
}
\affiliations {
    \textsuperscript{\rm 1}IBM, Boeblingen, Germany,\\ \textsuperscript{\rm 2}IBM Research, Dublin, Ireland\\
    aris.hofmann@ibm.com, ingevejs@ie.ibm.com, 
    dhaval.vinodbhai.salwala@ibm.com,
    elizabeth.daly@ie.ibm.com
}

\begin{document}
\maketitle
\begin{abstract}
We present Auto-BenchmarkCard, a workflow for generating validated descriptions of AI benchmarks. Benchmark documentation is often incomplete or inconsistent, making it difficult to interpret and compare benchmarks across tasks or domains. Auto-BenchmarkCard addresses this gap by combining multi-agent data extraction from heterogeneous sources (e.g., Hugging Face, Unitxt, academic papers) with LLM-driven synthesis. A validation phase evaluates factual accuracy through atomic entailment scoring using the FactReasoner tool. This workflow has the potential to promote transparency, comparability, and reusability in AI benchmark reporting, enabling researchers and practitioners to better navigate and evaluate benchmark choices.
\end{abstract}


\section{Introduction}
Benchmarks are vital in AI for standardizing tasks, enabling model evaluation, tracking progress, and setting baseline expectations \cite{reuel2024betterbench}. Appropriate benchmarks systematically  detect, assess, and mitigate risks \cite{sokol2024benchmarkcards}. Unsuitable benchmarks may risk leaving failure modes undetected, leading to deployment with unverified, poorly understood behaviors.
Choosing appropriate benchmarks ensures models are evaluated on relevant tasks, avoiding inaccurate assessments and missed risks.

Benchmark documentation is often limited, requiring developers to parse source code or consult reference papers to understand the details of the benchmark. Recently, \cite{sokol2024benchmarkcards} proposed a standardized representation of benchmark metadata, drawing inspiration from existing standards such as model cards \cite{mitchell2019model} and dataset documentation frameworks like Croissant \cite{akhtar2024croissant}. 
Sokol’s framework defines key benchmark metadata such as “purpose,” “methodology,” and “risks” in order to support informed selection, clearer stakeholder communication, and better understanding of objectives and limitations. However, creating benchmark cards by hand is time- and labor-intensive, posing a significant barrier to widespread adoption across the community. A large-scale analysis of Model Cards shows frequent completion of “Model Description” and “Training Procedure,” but critical fields—evaluation, limitations, and risks—are often left blank \cite{modelcardstats2025}. Regarding risks, \cite{rao2025riskrag} found that only 14\% of AI model cards mention risks, and 96\% of those were identical. This imbalance highlights the practical challenges of achieving comprehensive documentation through manual effort alone. To mitigate this, we introduce an automated workflow aimed at generating BenchmarkCards. The system employs a multi-agent architecture to extract relevant information from heterogeneous sources, including Unitxt, Hugging Face repositories, and associated academic publications. 
Content is structured into a BenchmarkCard based on Sokol’s schema \cite{sokol2024benchmarkcards}  and validated for factual consistency against source data.

\section{System Overview}
\begin{figure*}[t]
  \centering
  \includegraphics[width=1\linewidth]{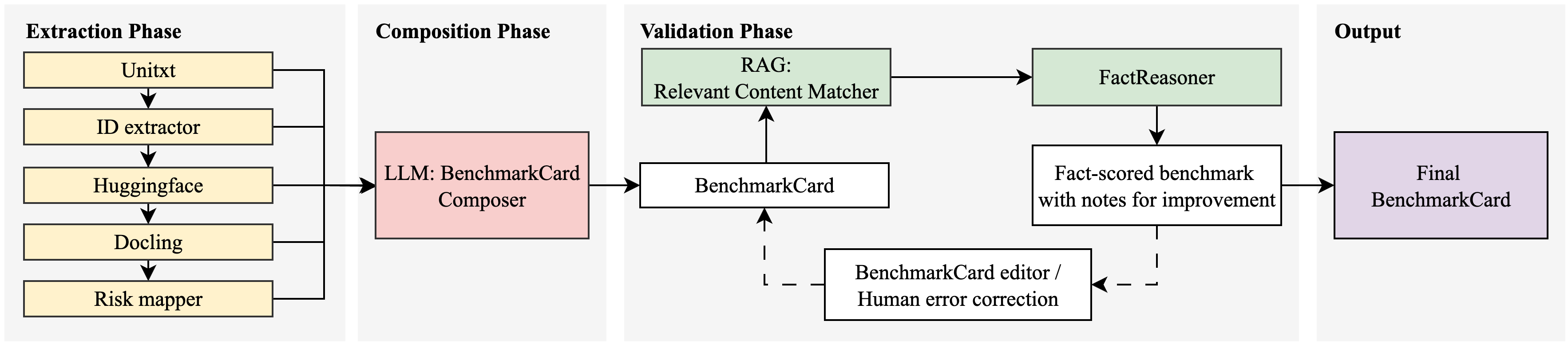}
  \caption{Auto-BenchmarkCard workflow overview}
  \label{fig:pipeline}
\end{figure*}

The workflow 
has three phases: \textit{Extraction}, \textit{Composition}, and \textit{Validation}, illustrated in Figure~\ref{fig:pipeline}. Access is provided via a Python CLI, and the system\footnote{\url{https://github.com/IBM/ai-atlas-nexus-demos/tree/main/auto-benchmarkcard}} is available in open source. 

\textbf{Extraction Phase:}
This phase collects structured benchmark data from multiple sources using modular custom agent tools. The implementation currently supports Unitxt but can be adapted for other standards like lm-eval-harness. Users begin by specifying a benchmark identifier for the \textit{Unitxt Tool}, built upon the Unitxt library \cite{bandel2024unitxt}, which searches its catalog and retrieves the corresponding \textit{UnitxtCard}. The retrieved card is then parsed to identify cited materials (e.g., metrics, templates) and retrieve related supplementary cards, with the result returned in JSON format. Next, the \textit{Extractor Tool} extracts identifiers from the JSON such as the Hugging Face repository ID and the publication URL for subsequent processing. The \textit{Hugging Face Tool} then extracts metadata from the benchmark's repository. Finally, the \textit{Docling Tool} \cite{livathinos2025docling} processes the benchmark's associated research publication, converting it into machine-readable markdown format.

\textbf{Composition Phase:} The extracted data is passed to a large language model (LLM), which generates a complete BenchmarkCard by filling predefined sections such as purpose, methodology, and limitations. Once the initial card is generated, the system passes it to the \textit{Risk Atlas Nexus} framework \cite{bagehorn2025riskatlas}. The risk identifier component flags potential risks based on a structured risk taxonomy. These risks are incorporated into the BenchmarkCard, adding a governance-informed layer to the technical content.

\textbf{Validation Phase:}
Validation 
plays a critical role by introducing a structured approach to verifying the factual accuracy of the initial BenchmarkCard. To address consistency challenges, especially conflicting details, we use \textit{FactReasoner}, a probabilistic framework for assessing factual consistency via natural language inference \cite{marinescu2025factreasoner}. We extend the package with custom components for atomization and context retrieval, specifically adapted to the BenchmarkCard format. The validation process begins by breaking down the BenchmarkCard into atomic statements: small, self-contained units of meaning that can each be checked for accuracy. This step 
is performed using an LLM with a prompt tailored to the structure and content of BenchmarkCards. In contrast to generic atomization, this approach ensures that the resulting statements are not only 
minimal but also 
and explicitly designed to be fact-checkable.

To assess each statement’s validity, we reference 
content from the Extraction Phase. 
The extracted metadata can be extensive, so evaluation by comparing each atomic statement against the full knowledge base 
is inefficient. To solve this, we index all extracted metadata used to compose the BenchmarkCard in a vector database. The retrieval module combines sparse keyword search and dense vector similarity to retrieve the relevant evidence for each atomic statement from the indexed knowledge documents. Retrieved chunks are 
graded and re-ranked by an LLM based on their relevance to the atomic statement. The paired atomic statements and their corresponding retrieved evidence are then passed to \textit{FactReasoner}, which assigns an entailment score between 0 and 1. A score close to 1 indicates that the statement is strongly supported by the source material (i.e., factually correct), a score near 0 indicates contradiction, and a score around 0.5 reflects a neutral or unverifiable claim. Atomic statements with low entailment scores are flagged for potential correction. Two remediation strategies are available: Automated Revision and Human-in-the-Loop Correction. \textit{Automated Revision:} Flagged sections are passed to an LLM along with their relevant 
context, allowing for targeted regeneration of specific BenchmarkCard fields with increased accuracy compared to the initial generation step, which relied on the full, unfiltered output of the Extraction Phase. \textit{Human-in-the-Loop Correction:} 
Flagged fields are routed for manual review and correction by human annotators.

This process results in a BenchmarkCard that has been automatically generated and validated for factual correctness, with the final card and workflow inputs/outputs provided as JSON.

\subsection{Limitations} 

The workflow has several limitations. First, performance is constrained by the completeness and quality of the extracted input. If insufficient data is retrieved from sources such as Hugging Face, 
the LLM may struggle to generate an 
accurate BenchmarkCard due to gaps in documentation. Second, factual correctness does not guarantee comprehensiveness or practical relevance. While the validation process ensures that each statement is grounded in evidence, it does not assess whether the content sufficiently covers all important aspects of the benchmark. Comprehensiveness refers to the extent to which the generated output addresses all relevant information, ensuring that no critical detail is omitted. In contrast, factuality only measures whether a given claim is accurate based on the available evidence, without introducing hallucinated content. This distinction becomes particularly important in cases where the generated content, though factually accurate, overlooks more central or representative information. For example, given a context stating “The main languages of the benchmark are English and Spanish, but some questions also address Portuguese,” a generated field listing only “Portuguese” as the benchmark language would pass a factuality check but fail in terms of comprehensiveness. It highlights a key limitation: accurate statements may still be misleading if they omit 
relevant information. 
To address this, one direction for future work 
would be to implement a dedicated evaluation step for comprehensiveness. 

\subsection{Outlook}

Our workflow produces a BenchmarkCard that is 
based on information extracted from multiple sources. Its factual alignment with the underlying data is assessed, and fields with low alignment scores are 
flagged for human intervention or revised by an LLM. While the workflow is specifically designed for BenchmarkCard generation, its modular architecture is 
applicable to similar tasks involving structured data extraction, LLM-based generation, and factuality validation. As such, it can be adapted to a wide range of use cases in automated documentation, summarization, and metadata synthesis across AI governance and other domains.

\bibliography{aaai2026}

\end{document}